\documentclass[final,5p,twocolumn]{elsarticle}

\usepackage{graphicx}          
\usepackage[dvips]{epsfig}    

\usepackage{xcolor}

\usepackage{amssymb}
\usepackage{amsthm}

\usepackage{amsmath}          

\usepackage{algorithmic}

\journal{Control Engineering Practice}

\begin{document}

\begin{frontmatter}

\title{Motion control with optimal nonlinear damping: from theory to experiment} 

\author[label1]{Michael Ruderman}
\ead{michael.ruderman@uia.no}


\address[label1]{Faculty of Engineering and Science,
University of Agder, Norway}

\begin{abstract}                          
Optimal nonlinear damping control was recently introduced for the
second-order SISO systems, showing some advantages over a
classical PD feedback controller. This paper summarizes the main
theoretical developments and properties of the optimal nonlinear
damping controller and demonstrates, for the first time, its
practical experimental evaluation. An extended analysis and
application to more realistic (than solely the double-integrator)
motion systems are also given in the theoretical part of the
paper. As comparative linear feedback controller, a PD one is
taken, with the single tunable gain and direct compensation of the
plant time constant. The second, namely experimental, part of the
paper includes the voice-coil drive system with relatively high
level of the process and measurement noise, for which the standard
linear model is first identified in frequency domain. The linear
approximation by two-parameters model forms the basis for
designing the PD reference controller, which fixed feedback gain
is the same as for the optimal nonlinear damping control. A robust
sliding-mode based differentiator is used in both controllers for
a reliable velocity estimation required for the feedback. The
reference PD and the proposed optimal nonlinear damping
controller, both with the same single design parameter, are
compared experimentally with respect to trajectory tracking and
disturbance rejection.
\end{abstract}


\end{frontmatter}

\newtheorem{thm}{Theorem}
\newtheorem{lem}[thm]{Lemma}
\newtheorem{clr}{Corollary}
\newdefinition{rmk}{Remark}
\newproof{pf}{Proof}

\section{INTRODUCTION}
\label{sec:1}

For the second-order systems it is understood that a linear
feedback control, see e.g. \cite{franklin2015} for basics, has
certain limitations in shaping the transient dynamics and
therefore the asymptotic convergence of the controlled output of
interest. In a standard state-space form
\begin{equation}\label{eq:1:1}
[\dot{x}_1,\dot{x}_2]^T =A \cdot [x_1,x_2]^T =
\left[%
\begin{array}{cc}
  0 & 1 \\
  -k & -d \\
\end{array}%
\right] \cdot[x_1,x_2]^T,
\end{equation}
the system matrix $A$ can be arbitrary shaped as Hurwitz via the
state-feedback coefficients $k,d>0$. Needless to dive into detail
that both coefficients can accommodate the own system dynamics as
well as the control gains. It is also worth recalling that a
standard linear proportional-derivative (PD) control, which is
sufficient for the unperturbed second-order systems, can be easily
integrated into the state-space form \eqref{eq:1:1}. Assuming the
output feedback gain $k$ is fixed by some dedicated control
specification or policy, like for example control saturations or
measurement noise, one can assign the linear damping term by
solving the associated characteristic polynomial
\begin{equation}\label{eq:1:2}
s^2+ds+k=(s+\lambda)^2
\end{equation}
with respect to $d$. Here the real double pole at $-\lambda$ is
optimal in terms of the damping (usually referred to as critical
damping). Namely, it shapes the control system in such a way that
it has neither a dominant (and thus slower) pole if $d>2\lambda$,
nor it oscillates transiently if $d<2\lambda$. It follows that the
linearly damped second-order dynamics of a feedback control system
of the type \eqref{eq:1:1} cannot perform any better convergence,
in the sense of an optimal damping rate, than that provided by the
real double pole. Few counterexamples can be found, like for
instance one of the optimal damping ratio for the linear
second-order systems which, however, requires the system damping
to be switched as a function of the system state
\cite{shahruz1992}. Also a comparative evaluation of different
controllers \cite{rao2001naive}, benchmarking for simplest
second-order system of a double integrator, can be mentioned here
as an associated reference.

The motion control systems often deal with the second-order
dynamics, in which the relative displacement $x_1$ and its rate
$x_2$ (i.e. velocity) of the moving tools and loads (in a more
general sense) are the state variables in focus. Practical
examples can range from the accurate micro- and nano-positioning
\cite{iwasaki2012,heertjes2015} to the standard robotic
manipulators \cite{anderson1988,dietrich2021}, just exemplary
referring to more former and more recent developments. Already in
the earlier works on the control in robotics, see e.g.
\cite{tomei1991}, it was recognized that a simple PD feedback
control is sufficient for regulation, once the main system
nonlinearities are compensated by either inverse dynamics control
or torque feed-forwarding. However, a certain temptation of
incorporating also nonlinear damping into the feedback of the
second-order systems, with the aim of improving the stabilization
and convergence properties, was occasionally made. This was
denoted, again in context of robotics, by the so-called nonlinear
proportional-derivative controllers, see e.g.
\cite{HolerPDnonlin95,kelly1996}. Recently, an optimal nonlinear
damping (OND) control, as combination with the standard
proportional output feedback, was proposed in \cite{ruderman2021a}
for the unperturbed second-order systems. This forms the basis of
the present work.

In this paper, we provide a practice oriented transition from the
theory to experiments for the OND control applied to the motion
control tasks. For the sake of a fair comparison, we also design a
robust PD feedback controller, which serves as a reference one,
and we stress that both controllers have only one and the same
tunable parameter -- the overall control gain. Since steering the
residual control errors towards zero is not our prime focus here,
it is explicitly emphasized that no additional integral control
actions are considered, so that a fair comparison to a standard
PID feedback control is not made. At this point, it is worth
noting that extension of the optimal nonlinear damping control by
an (eventually) nonlinear integral action might be an interesting
future research that requires further fundamental and extensive
investigations. Moreover one should state that the developed OND
control is suitable for the SISO systems, while a potential MIMO
extension is also subject to the future works. The rest of the
paper is divided into two main parts, theoretical and
experimental, accommodated in sections \ref{sec:2} and
\ref{sec:3}, correspondingly. The main conclusions are drawn in
section \ref{sec:4} at the end of the paper. The theoretical
developments from section \ref{sec:2} were partially presented in
\cite{ruderman2021b}. To complete the introduction, the overall
contribution of the paper can be summarized as follows.
\begin{itemize}

    \item The optimal nonlinear damping control of the second-order systems,
    proposed in \cite{ruderman2021a,ruderman2021b}, is provided in
    a consolidated manner for practical motion control
    applications. It includes a regularization factor, which
    extends the non-singular trajectory solutions to the whole $\mathbb{R}^2$
    state-space of the motion variables.

    \item Further theoretical developments and adjustments, in
    relation to the damped and perturbed motion system dynamics,
    are included. The motion system dynamics, identifiable in frequency
    domain, is addressed in regard to the control parametrization
    and tuning.

    \item Experimental evaluation of the nonlinear damping
    control is shown, for the first time, on a real drive system
    with inherent measurement and process noise. Robust
    sliding-mode differentiator is used for the not measurable
    relative velocity required for the feedback control. In addition,
    the proposed optimal nonlinear damping control is compared
    experimentally with a robustly designed linear PD feedback
    control.
\end{itemize}

\section{THEORETICAL PART}
\label{sec:2}

In this section, which is the theoretical part, we will summarize
the OND control which was first introduced in \cite{ruderman2021a}
and later shown in \cite{ruderman2021b} to have the convergent
dynamics with an augmented regularization factor. The convergent
dynamics, see \cite{pavlov2004}, will be briefly recalled for
convenience of the reader. It will also be shown how to apply the
OND control to the motion systems which have additional
first-order time delay dynamics and are, eventually, perturbed.
For those classical motion plants, we will also discuss the design
of a reference PD control, which can be tuned by only one free
parameter, similar as the OND control. Such design complexity
renders both controllers well comparable in a benchmarking.

\subsection{Optimal nonlinear damping control}
\label{sec:2:sub:1}

\begin{figure}[!h]
\centering
\includegraphics[width=0.85\columnwidth]{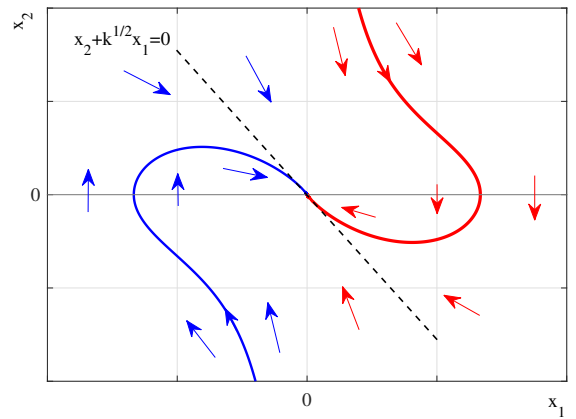}
\caption{Phase portrait of the closed-loop system
\eqref{eq:2:1:1}, \eqref{eq:2:1:2}.} \label{fig:2:1:1}
\end{figure}

\begin{figure}[!h]
\centering
\includegraphics[width=0.9\columnwidth]{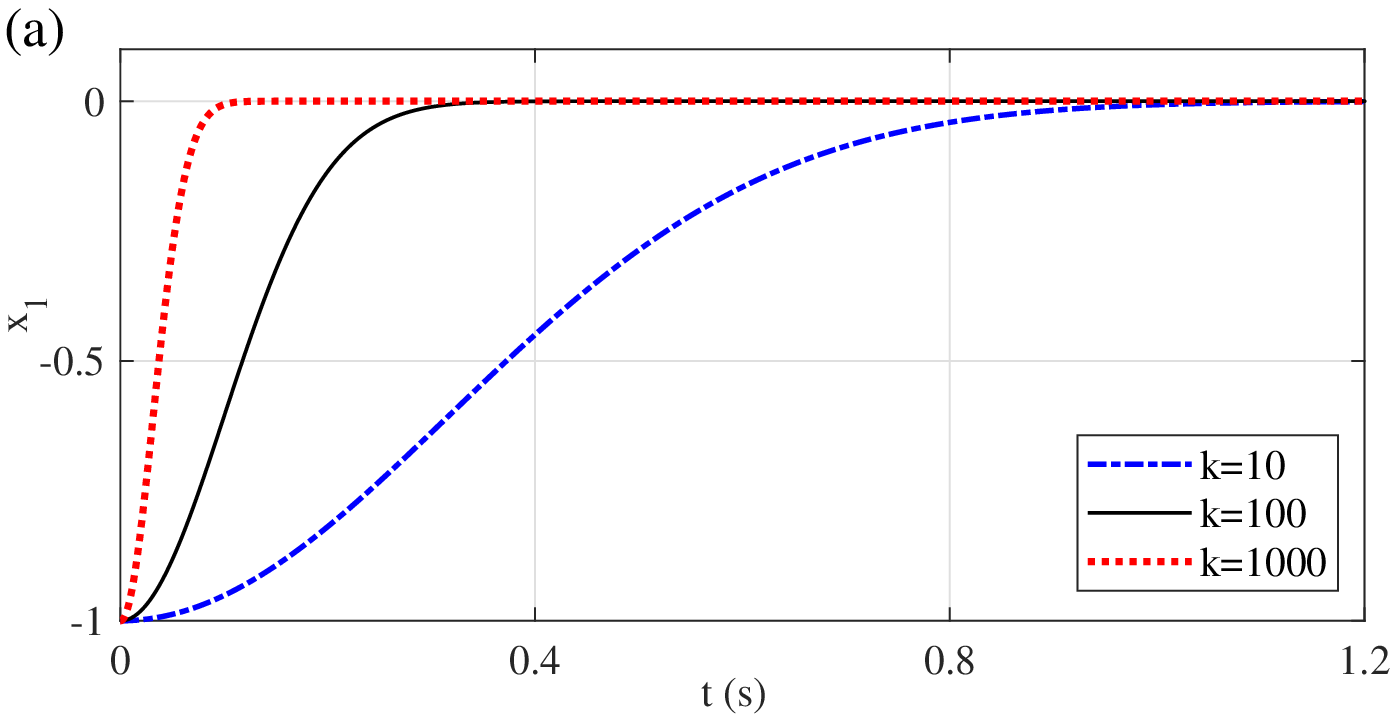}
\includegraphics[width=0.9\columnwidth]{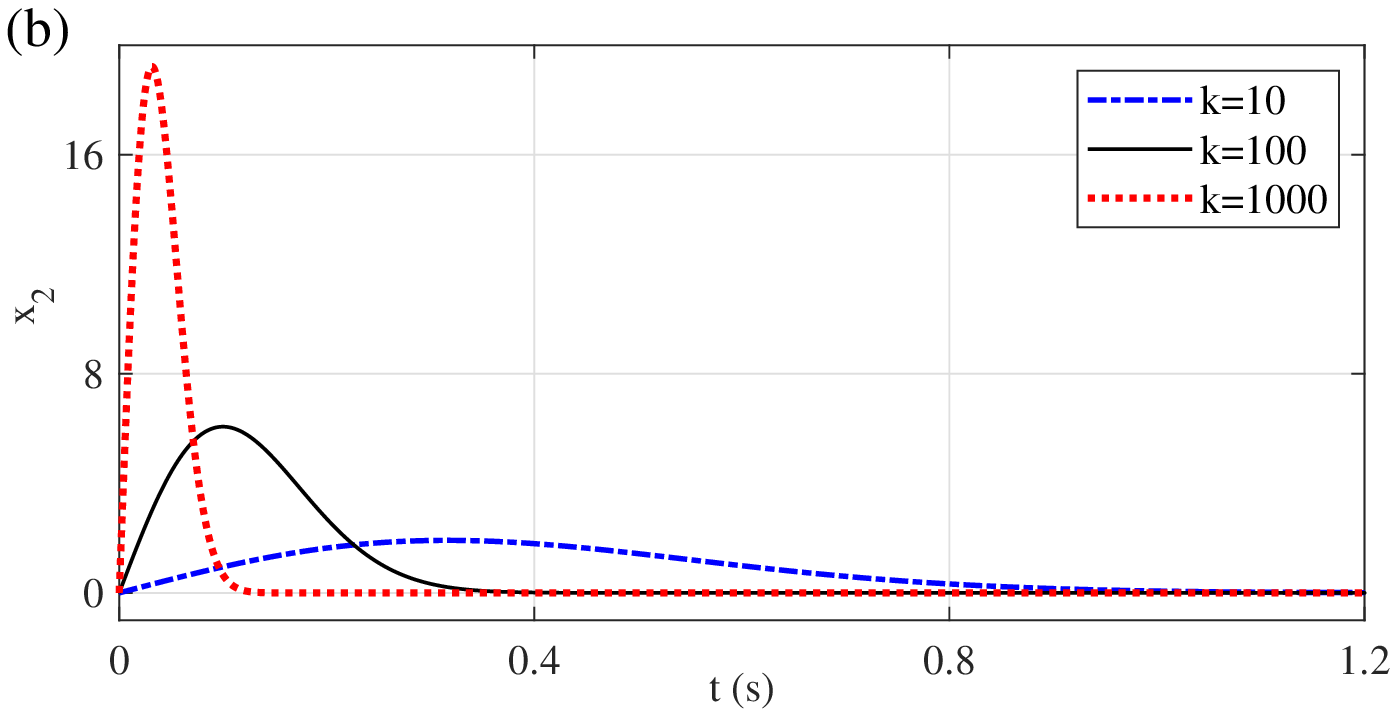}
\caption{State trajectories $x_1(t)$ in (a) and $x_2(t)$ in (b)
for the varying values of the control gain $k=
[10,\,100,\,1000]$.} \label{fig:2:1:2}
\end{figure}

The second-order closed-loop control system with an optimal
nonlinear damping is written as (cf. \cite{ruderman2021a})
\begin{eqnarray}
\label{eq:2:1:1}
  \dot{x}_1 &=&x_2,  \\
  \dot{x}_2 &=&-kx_1-x_2^2|x_1|^{-1} \mathrm{sign}(x_2), \label{eq:2:1:2}
\end{eqnarray}
where $k > 0$ is an arbitrary control design parameter. Note that
feedback control with the nonlinear damping term as in
\eqref{eq:2:1:2} was introduced first for unperturbed
double-integrator systems only. The system \eqref{eq:2:1:1},
\eqref{eq:2:1:2}, where $x_1$ is the output of interest, is
globally asymptotically stable and converges to the unique
equilibrium in the origin. This occurs: (i) along an attractor
\begin{equation}\label{eq:2:1:3}
x_2+ \sqrt{k} x_1=0
\end{equation}
in vicinity to the origin, and (ii) without crossing the
$x_2$-axis, see Figure \ref{fig:2:1:1}.
Note that the (ii)-nd property prevents singularity, which is
otherwise due to $x_1=0$ when $x_2(t) \neq 0$. It can be shown
that the closed-loop dynamics \eqref{eq:2:1:1}, \eqref{eq:2:1:2}
is always repulsing the state trajectories away from the $x_2=0$
axis, except from $(x_1,x_2)=\mathbf{0}$ equilibrium. The latter
is global and asymptotically attractive. Therefore, the admissible
set of the initial conditions for \eqref{eq:2:1:1},
\eqref{eq:2:1:2} is $X_0(t=0) = \bigl\{ (x_1,x_2) \in \mathbb{R}^2
\: |  \:  x_1 \in \mathbb{R}^{*} \bigr\}$, where $\mathbb{R}^{*}$
is the set of real numbers without null. The OND control does not
requires any gain (i.e tuning) parameter for the nonlinear damping
term, and the single output feedback gain $k$ is scaling the
transient response of both dynamic state trajectories, as
exemplary shown in Figure \ref{fig:2:1:2}. It is also worth
recalling that the closed-loop control system \eqref{eq:2:1:1},
\eqref{eq:2:1:2} allows for a bounded control action $|\dot{x}_2|
< S$, with $S = \mathrm{const}
> 0$. Such saturated control action, especially relevant for practical
applications, affects neither stability nor convergence properties
of the state trajectories, as shown in \cite{ruderman2021a}.

In order to allow for the state trajectories in the whole
$(x_1,x_2) \in \mathbb{R}^2$ state-space and, therefore, to avoid
singularity when crossing the $x_2$-axis outside the origin, a
regularization term $0 < \mu \ll k$ was later introduced in
\cite{ruderman2021b}. Moreover, the OND control was extended for
tracking the differentiable (at least once) reference trajectories
$r(t) \in \mathcal{C}^1$. For such reference signals, the OND
control performance is guaranteed for $\ddot{r}(t) = 0$, which can
be seen as steady-state for the motion control. For any
$\dot{r}(t) \neq \mathrm{const}$ finite-time (i.e. $t < T$)
transient phase trajectory, the OND control becomes temporary
perturbed. Then, i.e. for $t > T$, the OND control converges
according to the $(e_1,e_2)(t)$ error dynamics, where the output
error state is $e_1 = x_1-r$ and its time derivative is $\dot{e}_1
\equiv e_2 = x_2-\dot{r}$, respectively. The state error dynamics
of the regularized OND control, which is applied to the
double-integrator system, reads
\begin{eqnarray}
\label{eq:2:1:4}
  \dot{e}_1 &=&e_2,  \\
  \dot{e}_2 &=&-k e_1 - \frac{|e_2| \, e_2}{|e_1| + \mu}. \label{eq:2:1:5}
\end{eqnarray}
Note that the regularization term $\mu$ does not act as an
additional control gain, to be designed, but prevents singularity
in solutions of the closed-loop system \eqref{eq:2:1:1},
\eqref{eq:2:1:2}. When assuming a quadratic Lyapunov function
candidate
\begin{equation}\label{eq:2:1:6}
V = \frac{1}{2} k e_1^2 + \frac{1}{2} e_2^2,
\end{equation}
which represents the total energy level (i.e. potential energy of
the feedback control plus kinetic energy of the relative motion),
its time derivative results in
\begin{equation}\label{eq:2:1:7}
\frac{d}{dt}V = - \frac{|e_2|\,e_2^2}{|e_1| + \mu}.
\end{equation}
It can be recognized that while $V$ is positive definite and
radially unbounded, its time derivative \eqref{eq:2:1:7} is
negative definite $\forall \: e_2 \neq 0$ only. Applying the
standard invariance principle by LaSalle, one can easily show that
for $e_2 = 0$ outside the origin, the vector field $\dot{e}_2 = -k
e_1$ will always push a trajectory away from $e_2$-axis, where
$\dot{V}$ becomes negative definite. This proves the global
asymptotic stability of the unique equilibrium
$(e_1,e_2)=\mathbf{0}$.

\begin{figure}[!h]
\centering
\includegraphics[width=0.48\columnwidth]{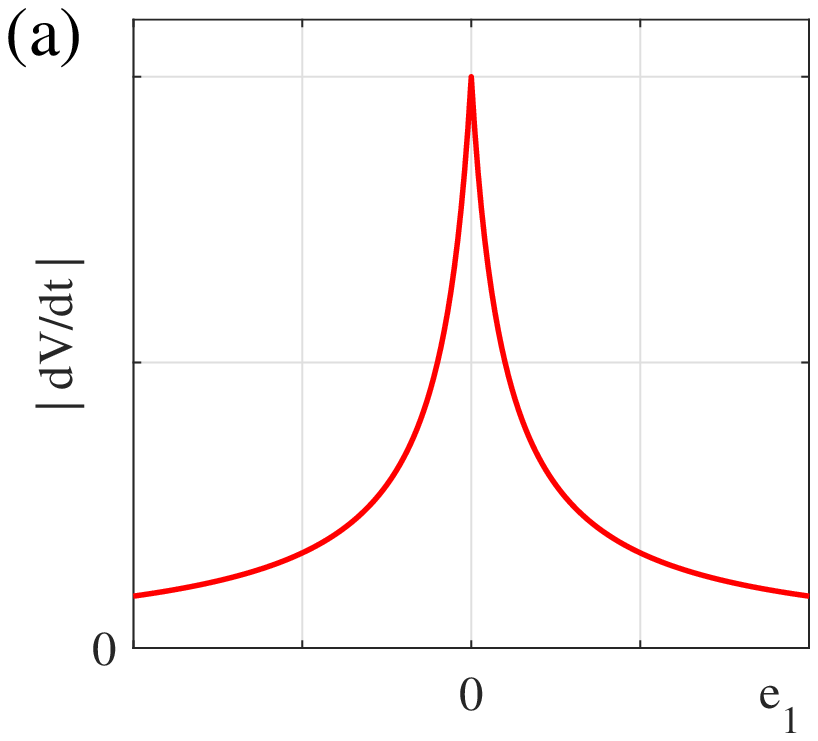}
\includegraphics[width=0.48\columnwidth]{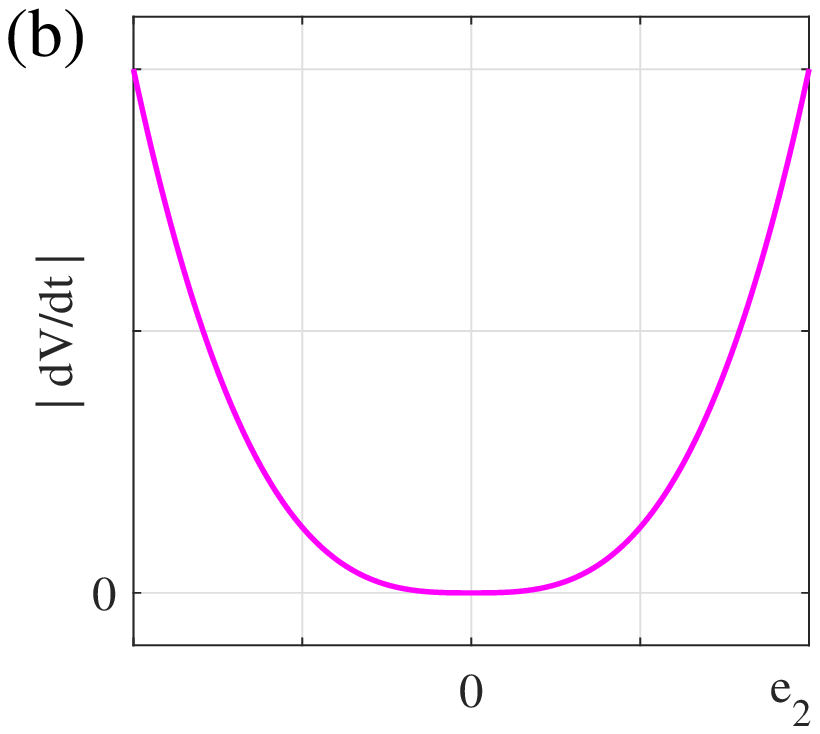}
\includegraphics[width=0.97\columnwidth]{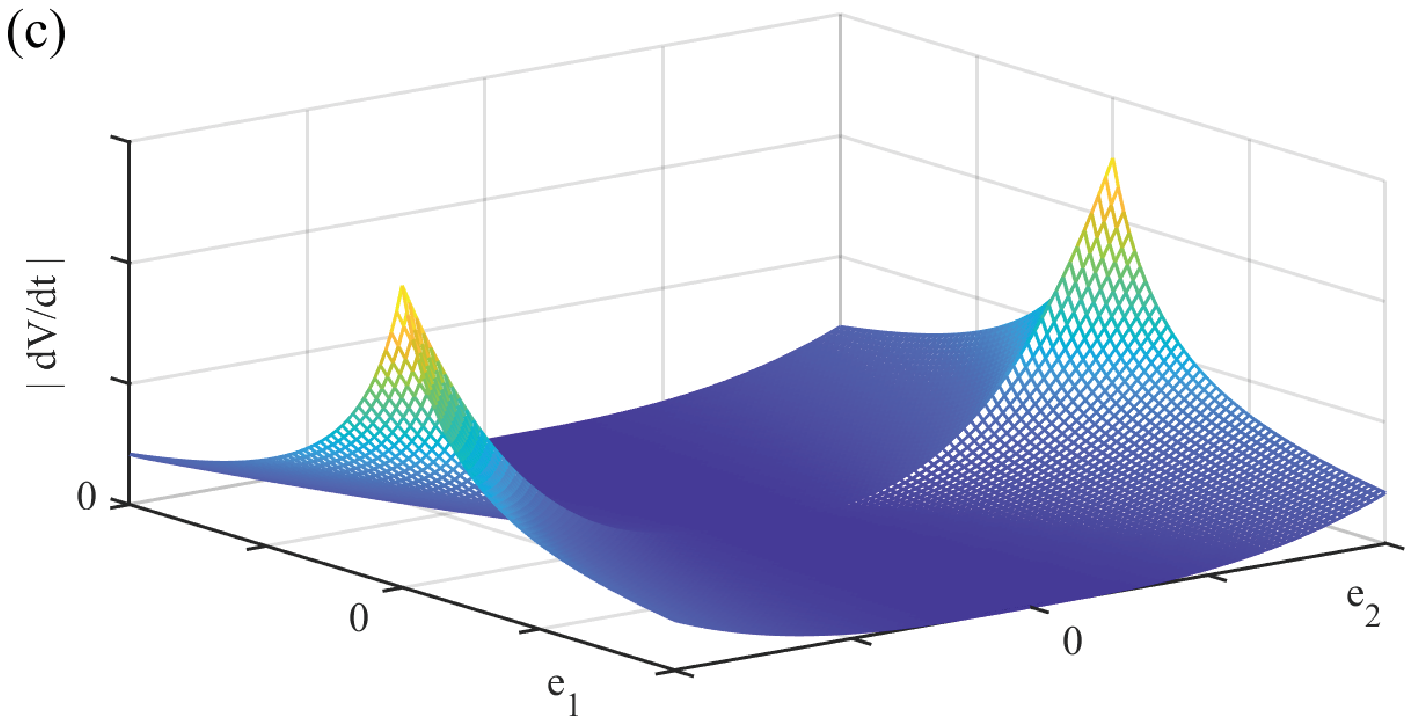}
\caption{Energy reduction rate $|\dot{V}|$ of the system
\eqref{eq:2:1:4}, \eqref{eq:2:1:5}: depending on $e_1$ in (a),
depending on $e_2$ in (b), and as overall error-states function
according to \eqref{eq:2:1:7} in (c).} \label{fig:2:1:3}
\end{figure}

A clarifying aspect of the nonlinear damping properties of
\eqref{eq:2:1:4}, \eqref{eq:2:1:5} highlights when analyzing the
rate at which the control system reduces its energy, based on
\eqref{eq:2:1:7}. From both projection of $|\dot{V}|$, shown in
Figure \ref{fig:2:1:3} (a) and (b), one can recognize that the
energy rate is hyperbolic in the error size, i.e. $\sim
|e_1|^{-1}$, and  cubic in the error rate, i.e. $\sim |e_2|^3$.
From Figure \ref{fig:2:1:3} (a), one can also recognized that the
regularization term $\mu$ prevents an infinite energy rate and,
thus, ensures a finite control action as $|e_1| \rightarrow 0$. At
the same time, a hyperbolic energy rate allows accelerating the
convergence as $|e_1| \rightarrow 0$. On the other hand, it is the
cubic dependency from the error rate which enables the OND control
reacts faster to the error dynamics, cf. Figure \ref{fig:2:1:3}
(b). This property is especially relevant for non-steady
trajectory phases, i.e. $\ddot{r}(t) \neq 0$, or when
perturbations provoke a fast growth of the $|e_2|$-value. The
overall landscape of the energy dissipation rate, see Figure
\ref{fig:2:1:3} (c), discloses that this is lower in magnitude
larger the control error norm $\| e_1,e_2\|$ is. For decreasing
$\| e_1,e_2\|(t)$, and that when $|e_1| \rightarrow 0$, the
$|\dot{V}|$ is largely growing, thus, allowing for faster
decelerations and convergence of the controlled motion in vicinity
to the reference trajectory.

\begin{figure}[!h]
\centering
\includegraphics[width=0.9\columnwidth]{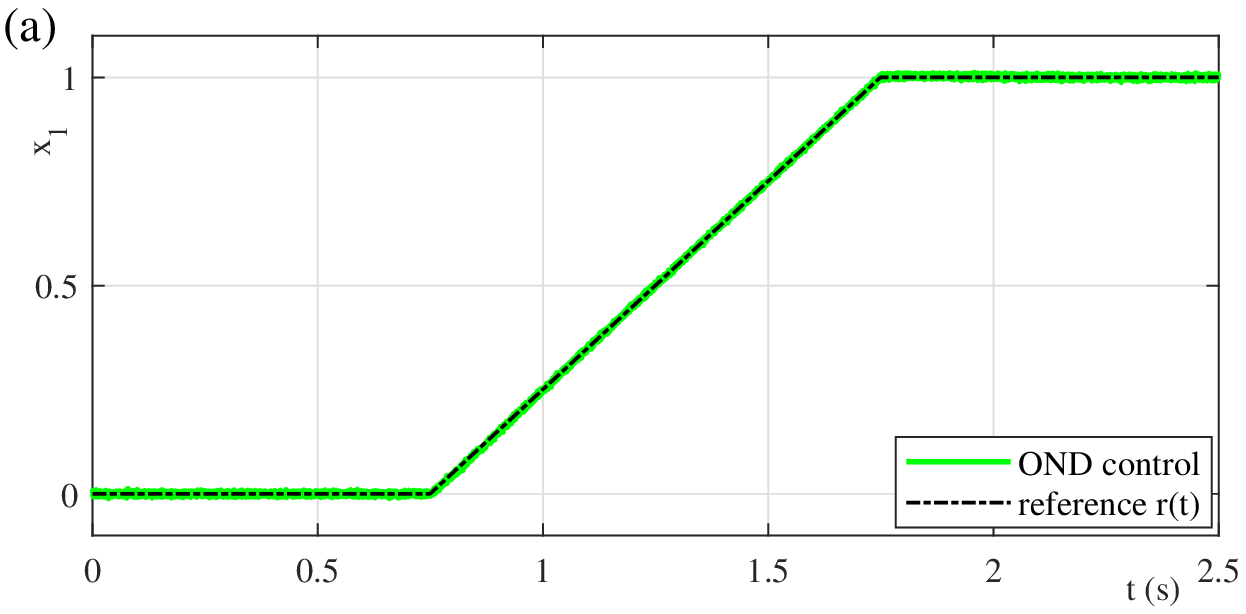}
\includegraphics[width=0.9\columnwidth]{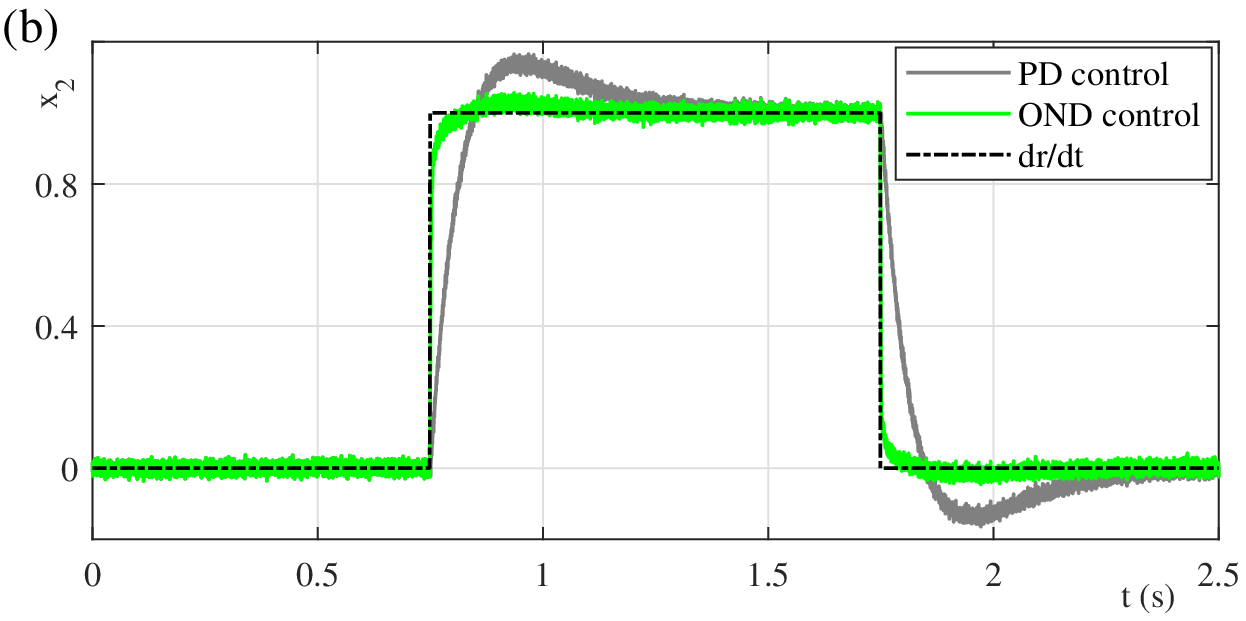}
\caption{Reference trajectory versus OND controlled output
position in (a), and comparison of the corresponding velocity for
OND and PD control in (b). Both controllers have the same feedback
gain factor $k=100$, while $\mu=0.0001$ is assumed for OND
control.} \label{fig:2:1:4}
\end{figure}

For better interpreting transient performance of the OND control,
let us compare it with the standard linear PD feedback control,
for which the error dynamics \eqref{eq:2:1:5} transforms to
$\dot{e}_2 = -k e_1 - k \, \tau e_2$, where $\tau$ is the time
constant parameter. Obviously, the $k$ and $\tau$ parameters can
be assigned so that the linear closed-loop dynamics is critically
damped, i.e. has a negative double real pole at the desired
location. For the following numerical example, let us assume
$k=100$ and $\tau=0.2$, resulting in the double real pole at
$-10$. Note that we assume the same $k=100$ for the OND control,
given by \eqref{eq:2:1:4}, \eqref{eq:2:1:5}, while $\mu=0.0001 \ll
k$ is assigned. The piecewise linear trajectory $r(t)$, that is
typical for motion control tasks, is exemplary shown in Figure
\ref{fig:2:1:4} (a), together with the OND controlled output
trajectory. The difference in transient response between the OND
control and critically damped PD control, both having the same
feedback gain factor, is best visible in the $x_2(t)$ trajectories
shown in Figure \ref{fig:2:1:4} (b).

\begin{figure}[!h]
\centering
\includegraphics[width=0.9\columnwidth]{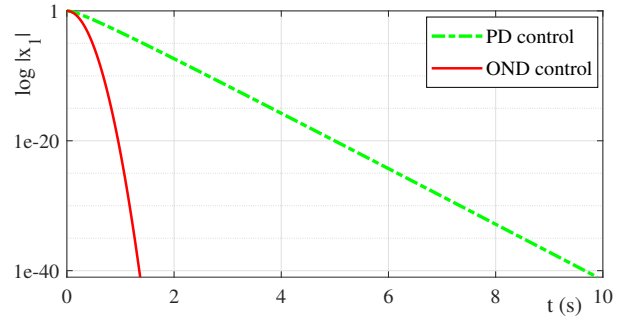}
\caption{Convergence of the controlled output value of the PD and
OND controllers (logarithmic absolute value).} \label{fig:2:1:5}
\end{figure}

The convergence properties of both controllers become even more
evident when assuming $r=0$ and $x_1(0) \neq 0$ and comparing the
$x_1(t)$ trajectories plotted on the logarithmic scale, see Figure
\ref{fig:2:1:5}. While the linear PD control shows the (expected)
linear-shaped convergence on the logarithmic scale, the OND
control discloses a hyper-exponential (e.g. quadratic on the
logarithmic scale) convergence of $|x_1|(t)$. It is easy to
recognize that the difference and, therefore, advantage of the OND
control becomes more considerable, higher the control accuracy
and, correspondingly, lower residual error $e_1(t) \, | \, _{t
\rightarrow \infty}$ are required.

\subsection{Convergent dynamics}
\label{sec:2:sub:2}

Below, we briefly recall the main statements and properties of a
dynamic system to be convergent, according to
\cite{demidovich1967}, while for more details we refer to
\cite{pavlov2004}, and for the convergent system \eqref{eq:2:1:4},
\eqref{eq:2:1:5} to \cite{ruderman2021b}.

\emph{Definition 1.} \label{def:1} The system $\dot{x} = f(x,t)$
is said to be convergent if for all initial conditions $t_0 \in
\mathbb{R}$, $\bar{x}_0 \in \mathbb{R}^n$ there exists a solution
$\bar{x}(t) = x(t,t_0,\bar{x}_0)$ which satisfies:
\begin{enumerate}
    \item[(i)] $\bar{x}(t)$ is well-defined and bounded for all $t \in (-\infty,
    \infty)$;
    \item[(ii)]$\bar{x}(t)$ is globally asymptotically
    stable.
\end{enumerate}
Such solution $\bar{x}(t)$ is called a \emph{limit solution}, to
which all other solutions of the dynamic system converge as $t
\rightarrow \infty$.

\emph{Theorem 1.}\label{thm:1} Consider the system $\dot{x} =
f(x,t)$. Suppose, for some positive definite matrix $P=P^T > 0$
the matrix
\begin{equation}\label{eq:2:2:1}
J(x,t) := \frac{1}{2} \Bigl ( P \, \frac{\partial f}{\partial
x}(x,t) + \Bigl[ \frac{\partial f}{\partial x}(x,t) \Bigr]^T P
\Bigr)
\end{equation}
is negative definite uniformly in $(x,t) \in \mathbb{R}^n \times
\mathbb{R}$ and $|f(0,t)| \leq \mathrm{const} < +\infty$ for all
$t \in \mathbb{R}$. Then the system is convergent. The detailed
proof can be found in \cite{pavlov2004}.

Evaluating the Jacobian of $f(x,t)$ of the system
\eqref{eq:2:1:4}, \eqref{eq:2:1:5} with $x = [e_1,\,e_2]^T$ and
suggesting
\begin{equation}\label{eq:2:2:2}
P = \frac{1}{2} \, \left[%
\begin{array}{cc}
  k & 0 \\
  0 & 1 \\
\end{array}%
\right],
\end{equation}
which is the positive definite matrix, one can show that the
matrix $J(x,t)$, which is the solution of \eqref{eq:2:2:1}, is
negative definite and, hence, the Theorem 1 holds. For proving it,
we evaluate the matrix definiteness as
\begin{equation}\label{eq:2:2:3}
x^T  J(x,t) \, x = - \frac{3}{4} \, \frac{|e_2| \,e_2^2
\,\bigl(|e_1| + 2\mu \bigr)} { \bigl(e_1 + \mu \,
\mathrm{sign}(e_1) \bigr)^2} \, \leq 0 \quad \forall \; x\neq 0.
\end{equation}
Note that the obtained inequality \eqref{eq:2:2:3} proves only the
negative \emph{semi-definiteness} of $J(x,t)$, since $x^T J(x,t)
\, x = 0$ for $e_2 = 0  \, \wedge \, e_1 \neq 0$. It is however
possible to show that $[e_1,\,e_2] = 0$ is the unique limit
solution by evaluating the $\dot{e}_2$ dynamics at $e_2=0$.
Substituting $e_2=0$ into \eqref{eq:2:1:5} results in $\dot{e}_2 =
- k e_1$. It implies that $[e_1 \neq 0, \, e_2 = 0](t)$ cannot be
a limit (correspondingly steady-state) solution, since any
trajectory will be repulsed away from $e_2 = 0$ as long as $e_1
\neq 0$. Therefore, the closed-loop control system
\eqref{eq:2:1:4}, \eqref{eq:2:1:5} is uniformly convergent.
Consequently, the origin in the control error coordinates
$[e_1,e_2](t) = 0 \equiv \bar{x}$ is the unique limit solution for
$\forall \: t_0 < \tau < t$, independently of the initial
conditions $[e_1,e_2](t_0)$.

\subsection{Control extension for common motion systems}
\label{sec:2:sub:3}

The practical motion systems, associated with the controlled
drives, contain usually the additional damping dynamics and input
gain parameters, so that a system to be controlled is no longer
just the double integrator, cf. \eqref{eq:2:1:1},
\eqref{eq:2:1:2}. Without restoring force actions and, therefore,
having one free integrator (in terms of $\dot{x}_1 = x_2$), the
motion dynamics which is driven by the control input $u$ can be
most simply modeled by
\begin{equation}\label{eq:2:3:1}
\tau \dot{x}_2(t) + x_2(t) = K u(t).
\end{equation}
Here, the time constant $\tau = m \, \sigma^{-1}$ results from the
overall moving mass $m$ and linear (viscous) damping coefficient
$\sigma$, while $K$ is the overall input gain which converts the
available control channel into the generalized force quantity. The
latter is, correspondingly, actuating the drive system according
to the Newton's laws of motion. Obviously, such linear system
\eqref{eq:2:3:1} can be directly transformed into Laplace domain
and then identified (i.e. also in frequency domain). The latter
will result in determining only two free parameters, $\tau$ and
$K$, when using the input $u(t)$ and available output measurement,
either $x_1(t)$ or $x_2(t)$, cf. with the experimental part
provided in section \ref{sec:3}.

For motion systems, which are not the free double-integrator but
have dynamics of the form \eqref{eq:2:3:1}, the parametric scaling
of the OND control is required. This is in order the closed-loop
control system keeps the same stability and convergence properties
as for \eqref{eq:2:1:4}, \eqref{eq:2:1:5}. First, consider the
system plant \eqref{eq:2:3:1} without its linear damping term.
Substituting the OND control instead of $u$ results in
\begin{equation}\label{eq:2:3:2}
\dot{x}_2(t) = \frac{K}{\tau} \bigl( \mathcal{P} + \mathcal{D}
\bigr),
\end{equation}
where the proportional and damping control parts are abbreviated
(for the sake of clarity) by $\mathcal{P}$ and $\mathcal{D}$,
correspondingly, cf. \eqref{eq:2:1:1}, \eqref{eq:2:1:2}. Recall
that the proportional control part $\mathcal{P}= -k (x_1-r)$
allows for any positive gain values $k >0$, without affecting the
basic properties of OND control, cf.
\cite{ruderman2021a,ruderman2021b}. Therefore, no scaling of
$\mathcal{P}$ is required. At the same time, one can recognize
that an inverse gaining factor $\tau/K$ must be incorporated into
$\mathcal{D}$ for keeping the left- and right-hand side of
\eqref{eq:2:3:2} as balanced as in the original OND control, cf.
\eqref{eq:2:1:2}. Now, taking back into account the linear system
damping, which is scaled by $\tau^{-1}$ cf. with \eqref{eq:2:3:1},
one obtains the closed-loop dynamics
\begin{equation}\label{eq:2:3:3}
\dot{x}_2(t) + \frac{1}{\tau} x_2(t) = \frac{K}{\tau} \bigl(
\mathcal{P} + \mathcal{D} \bigr),
\end{equation}
with the scaled damping control part
\begin{equation}\label{eq:2:3:4}
\mathcal{D} = - \frac{\tau}{K} \, \frac{|e_2| \, e_2}{|e_1| +
\mu}.
\end{equation}
Through the applied scaling of the OND control in
\eqref{eq:2:3:4}, the inertial (on the left-hand side) and control
(on the right-hand side) terms in \eqref{eq:2:3:3} will represent
the nonlinear differential equation with the same convergence
properties as \eqref{eq:2:1:5}. However, the linear damping term
(on the left-hand side of \eqref{eq:2:3:3}) appears now as a
disturbing factor. This can be compensated by direct inclusion
into the control law, i.e. on the right-hand side of
\eqref{eq:2:3:3}. The resulted control law of the scaled OND with
an additional compensation of the system damping term is
\begin{equation}\label{eq:2:3:5}
u(t) = \mathcal{P} + \mathcal{D} + \frac{1}{K} x_2(t).
\end{equation}

Now, taking into account that the nominal motion dynamics
\eqref{eq:2:3:1} can be perturbed by some upper bounded
disturbance $|\xi| \leq \mathrm{const}$, the closed-loop behavior
of the plant \eqref{eq:2:3:1} with the control \eqref{eq:2:3:5}
results in
\begin{equation}\label{eq:2:3:6}
\tau \ddot{x}_1 + \tau \frac{|\dot{x}_1|\dot{x}_1}{|x_1| +\mu} + k
x_1 = \xi,
\end{equation}
when assuming $r=0$, for the sake of simplicity, and some non-zero
initial conditions $(x_1,x_2)(t) \neq 0$ for the sake of analysis.
It can be seen that the second-order nonlinear differential
equation \eqref{eq:2:3:6} remains asymptotically stable, cf.
sections \ref{sec:2:sub:1} and \ref{sec:2:sub:2}, while no longer
converging to zero equilibrium but to $x_1(t) \rightarrow \xi
k^{-1}$ at steady-state. This indicates how large the static
position control error is, just as in case of standard PD feedback
controllers when they are affected by the matched perturbations.
If the matched perturbation $\xi(t)$ appears dynamically, the
convergence of the OND control will always be faster than that of
a classical PD controller, cf. Figure \ref{fig:2:1:5}.

\subsection{Design of reference PD controller}
\label{sec:2:sub:4}

As a reference PD feedback controller, we assume the one which has
the following form
\begin{equation}\label{eq:2:4:1}
u(t) = \gamma \bigl(r(t) - x_1(t) \bigr) - \gamma \tau x_2(t),
\end{equation}
with the design gain factor $\gamma$ and given parameter $\tau$.
The latter compensates directly for the time constant of the
system plant, cf. \eqref{eq:2:3:1}, this way making the PD control
\eqref{eq:2:4:1} to a simply tunable one-parameter feedback
regulator. Note that solely the control error $e(t) \equiv r(t) -
x_1(t)$ is subject to the proportional control amplification,
while the differential control part with the total gaining by
$\gamma \tau$ is using the output velocity and not $\dot{e}(t)$.
This allows applying also the discontinuous reference signals,
like e.g. a step, for which $\dot{r}$ does not exist. For
analyzing optimality of parametrization of the PD control
\eqref{eq:2:4:1}, one can easily extend the right-hand-side of
\eqref{eq:2:4:1} by $\gamma \tau \dot{e}(t)$ and, after
substituting $u(t)$ into \eqref{eq:2:3:1}, obtain the open-loop
transfer function in Laplace domain as
\begin{equation}\label{eq:2:4:2}
x_1 s (\tau s + 1 ) = K \gamma \, e (\tau s + 1 ).
\end{equation}
Obviously, the pole-zero cancelation in \eqref{eq:2:4:2} converts
the open-loop transfer function into the simple integrator which
is amplified by the factor $K \gamma$. Transforming it back into
time domain and writing out the control error gives
\begin{equation}\label{eq:2:4:3}
\dot{x}_1(t) + K \gamma \, x_1(t) = K  \gamma \, r(t).
\end{equation}
This yields a principal first-order closed-loop dynamics which has
zero steady-state error and a time constant which is arbitrary
assignable through the control gain $\gamma$. In practical
applications, a motion system \eqref{eq:2:3:1} will have also some
neglected or parasitic (additional) dynamics at higher-frequencies
and, thus, a deteriorated phase response (in terms of a phase
margin) as implication. Therefore, the control gain $\gamma$ needs
to be assigned with respect to the resulted cross-over frequency
and the associated stability margins, cf. the practical part in
section \ref{sec:3:sub:3}.

\section{EXPERIMENTAL PART}
\label{sec:3}

\begin{figure}[!h]
\centering
\includegraphics[width=0.85\columnwidth]{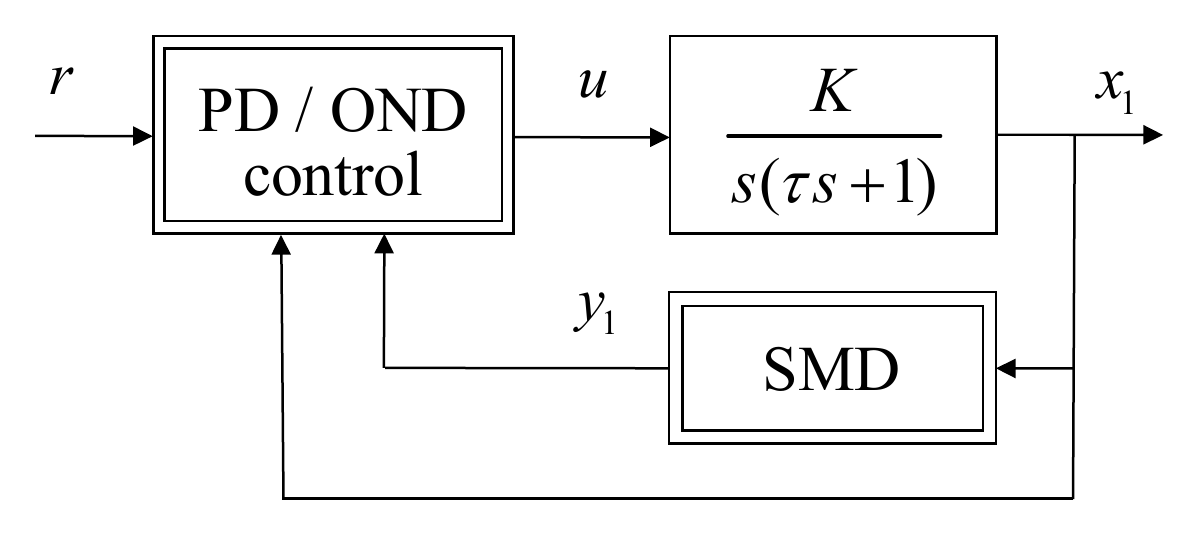}
\caption{Block diagram of the closed-loop control system.}
\label{fig:3:0}
\end{figure}

This section is dedicated to an experimental case study, showing
practical applicability and resulted performance of the OND
control. The block diagram of the entire closed-loop control
system, designed according to the section \ref{sec:2} and
evaluated experimentally as follows, is depicted in Figure
\ref{fig:3:0}. The second-order plant of a motion system includes
the free integrator and dynamics \eqref{eq:2:3:1},
Note that both controllers in use, i.e. either PD or OND one, are
served by the same input signals, -- the reference value $r(t)$,
the measured system output $x_1(t)$, and its derivative $y_1(t)$,
obtained by means of the SMD, cf. with section \ref{sec:3:sub:2}.

\subsection{Second-order motion system with voice-coil drive}
\label{sec:3:sub:1}

\begin{figure}[!h]
\centering
\includegraphics[width=0.5\columnwidth]{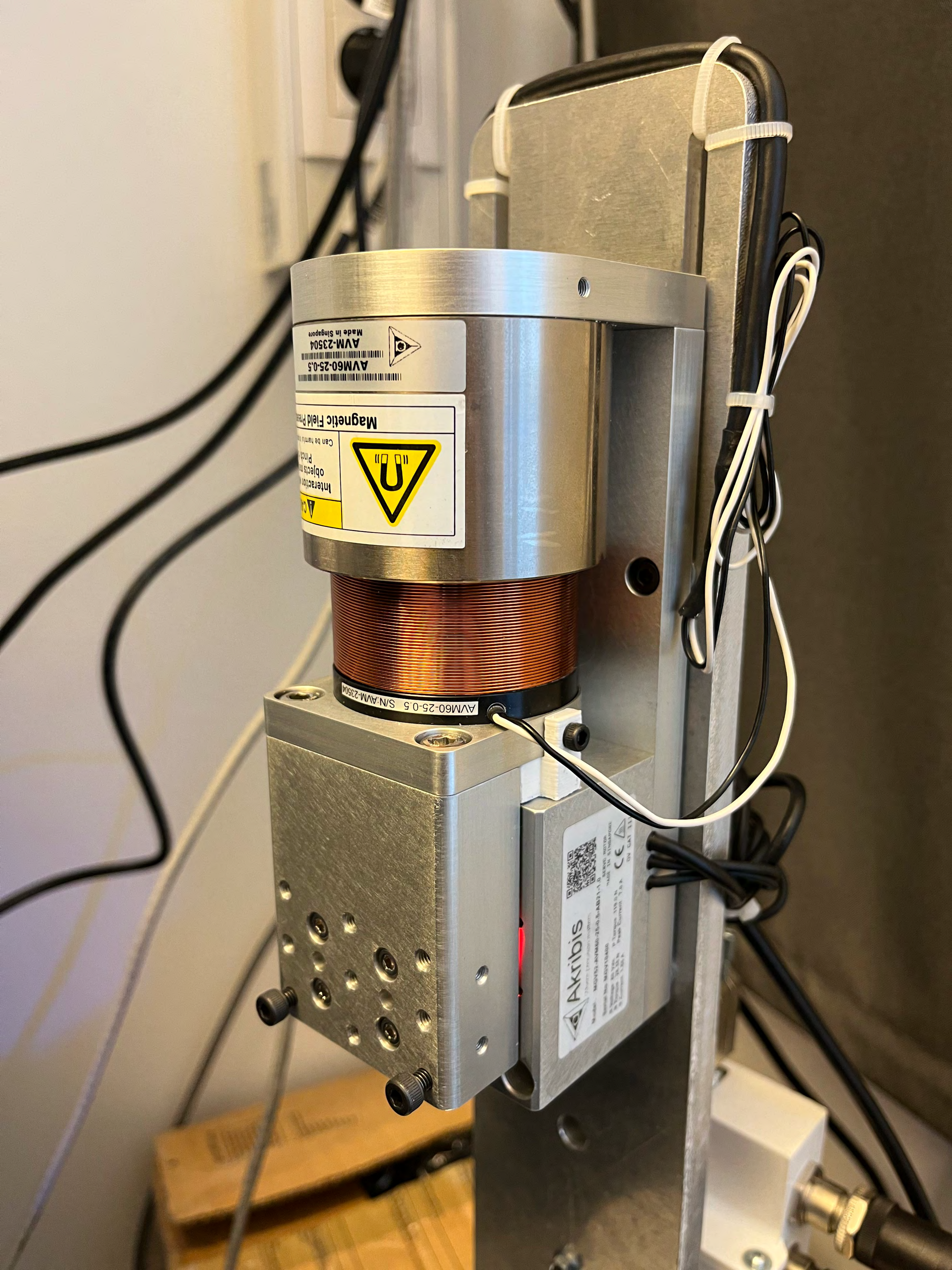}
\caption{Experimental setup of the voice-coil drive with 1DOF.}
\label{fig:3:1:1}
\end{figure}

The second-order motion system under investigation is the
voice-coil drive, shown in the laboratory setting in Figure
\ref{fig:3:1:1}. The electro-magnetically actuated voice-coil
motor has the total linear stroke about 20 mm, which is indirectly
measured by the contactless inductive displacement sensor with a
nominal repeatability of $\pm 12$ $\mu m$.
Due to a specific attachment of the moving rigid bar, which is
entering detection area of the contactless sensor, the effective
measurement range and, therefore, displacement operation range
$x_1$ of the drive is limited to about 13 mm only. The system
discloses a relatively large sensor and process noise. The former
is due to contactless sensing, while the latter is due to
additional parasitic dynamic by-effects which are not captured by
the second-order motion dynamics. The nominal electrical time
constant, which is however neglected when modeling the motion
dynamics is 1.2 msec. The real-time control board operates the
system with the set sampling rate of 10 kHz, while the available
control signal is the power-amplified voltage $U$ in the range
$[0,\,10]$ V. The overall voice-coil motor resistance is $R =
5.23$ Om, while the nominal motor force constant is $\Psi = 17.16$
N/A. The overall moving mass of the drive, determined by the scale
measurement of the parts and technical data-sheet of the drive, is
$m = 0.538$ kg.

The hardware specific properties of the voice-coil motor drive
require the following measures to be taken with the input and
output signal channels, so as to apply the feedback control
provided in section \ref{sec:2}. When neglecting the non-modeled
dynamics of electro-magnetic circuit, the input force constant is
$K_u = \Psi R^{-1}$. Note that $K_u$ factor has the N/V units,
since mapping the input voltage $U$ to the induced input motor
force $u$, and is linked to the overall input gain by $K = K_u
\sigma^{-1}$, cf. \eqref{eq:2:3:1}. The voice-coil motor drive in
its vertical arrangement, cf. Figure \ref{fig:3:1:1}, is subject
to the constant gravity force $mg$, where $g=9.8$ m/s$^2$ is the
gravitational acceleration constant. Furthermore, the stator-mover
configuration of the voice-coil motor gives rise the to the
periodic force ripples, which act as a 'magnetic stiction' force
when the motor starts to move. In order to overcome it, a square
pulse jitter signal $U_j$ of a low amplitude 0.2 V and high
frequency 450 rad/sec is used. The suitable amplitude and
frequency are found (experimentally) so that to not induce an
effective motion above the measurement noise of the displacement
sensor, on the one hand. On the other hand, the oscillating jitter
signal should be sufficient to overcome the periodic
position-dependent force ripples. The determined jitter frequency
is also clearly above the bandwidth and, thus, cutoff frequency of
the designed feedback controllers. By incorporating the above
mentioned jitter and gravity compensation signals, the overall
control voltage becomes
\begin{equation}\label{eq:3:1:1}
U(t) = U_j(t) + \frac{mg}{K_u} + u(t),
\end{equation}
where the feedback control signal $u(t)$ takes explicitly into
account the input gain $K_u$, cf. section \ref{sec:2:sub:3}.

\subsection{Robust sliding-mode based differentiator}
\label{sec:3:sub:2}

Since both the OND and the reference PD controllers require the
unavailable $x_2(t)$-signal in feedback, its trustfully estimated
value must be obtained from the measured $x_1(t)$. Applying a
real-time discrete-time differentiation of the $x_1(t)$-signal
does not provide an operational and robust solution for feedback
of the velocity feedback. This is due to the measured position
(cf. section \ref{sec:3:sub:1} above) contains the broadband
components of process and measurement noise and, therefore, does
not necessarily hold a sufficient SNR (signal-to-noise ratio). The
differentiated signals from the position measurement require
mostly a low-pass filtering which inserts an additional phase lag
into the control loop. This can largely restrict the achievable
bandwidth of the control system and, generally, reduce the
closed-loop performance at higher angular frequencies.

Robust differentiators \cite{levant1998}, which are based on the
sliding-mode principles see e.g. \cite{shtessel2014}, provide an
alternative for obtaining the fast estimation of relative velocity
in real-time. Here the remarkable features are an insensitivity to
the bounded noise (provided the Lipschitz constant of n-th
time-derivative is available) and a finite-time convergence. The
latter makes a robust sliding-mode based differentiator
theoretically free of a phase lag which is, otherwise, unavoidable
for the low-pass filtering. Assuming the estimation error of the
robust sliding-mode based differentiator (further as SMD) is
$\varepsilon(t) = y_0(t) - x_1(t)$, the second-order SMD, cf.
\cite{moreno2012}, is given by
\begin{eqnarray}
\label{eq:3:1:2}
  \dot{y}_0 &=& -\kappa_0 |\varepsilon|^{2/3} \mathrm{sign}(\varepsilon) + y_1, \\
\label{eq:3:1:3}
  \dot{y}_1 &=& -\kappa_1 |\varepsilon|^{1/3} \mathrm{sign}(\varepsilon) + y_2, \\
\label{eq:3:1:4}
  \dot{y}_2 &=& -\kappa_2 \mathrm{sign}(\varepsilon).
\end{eqnarray}
Note that the second-order (and not first-order) SMD is
purposefully assumed here, in order to obtain a smoother estimate
$y_1(t)$ of the relative velocity. Recall that the robust
second-order SMD provides $y_0(t) = x_1(t)$, $y_1(t) =
\dot{x}_1(t)$, $y_2(t) = \ddot{x}_1(t)$, for all $t > t_c$, where
$t_c$ is a finite convergence time. Also important to emphasize is
that the Lipschitz constant $L$ of $\ddot{x}_1$ needs to be known
and, thus, the upper bound of the highest derivative
$|\dddot{x}_1| \leq L$. While different parametrization approaches
for $\kappa_n$, all taking into account $L$, exist in the HOSM
(high-order sliding mode) literature, the parametrization provided
in \cite{reichhartinger2017} is used in the following. The scaling
factor $\rho$ is used so that $\kappa_n = k_n \rho^{n+1}$, where
$\rho^{n+1}$ (in our case $\rho^3$) corresponds to the Lipschitz
constant $L$ of the highest derivative ${x}_1^{(n)}$. The
coefficients $k_{0,1,2} = \{3.1, \, 3.2, \, 1.1 \}$ of the
second-order SMD are used according to \cite{reichhartinger2017},
while $L$ and, correspondingly, $\rho$ remain unknown for the
given experimental system.
\begin{figure}[!h]
\centering
\includegraphics[width=0.95\columnwidth]{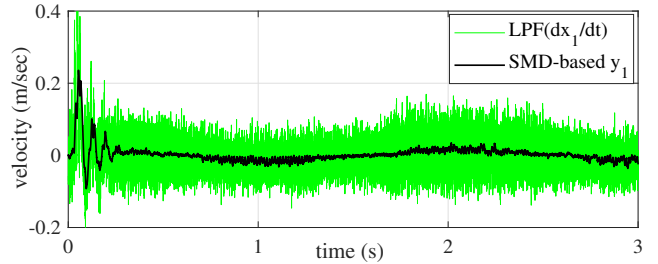}
\caption{SMD based estimation $y_1(t)$ of the relative velocity
versus low-pass filtered discrete time derivative of the measured
$x_1(t)$.} \label{fig:3:2:1}
\end{figure}
Therefore, the applied scaling factor $\rho=8$ is tuned
experimentally so that the estimate $y_1(t)$ is sufficiently
accurate with respect to $\dot{x}_1(t)$. Note that the latter can
be computed (for tuning purposes) as a smooth and noise-free
signal, since the measured $x_1(t)$ fits with $a_i \sin (\omega_i
t)$ for the driven constant amplitude $a_i$ and frequency
$\omega_i$. The driven and measured $x_1(t)$ with two limiting
angular frequencies $\omega_i = \{0.1,\,10\}$ rad/sec where used
for tuning the $\rho$-parameter. For highlighting the resulted SMD
performance, the $y_1(t)$-estimation of the relative velocity is
exemplary compared in Figure \ref{fig:3:2:1} with the
discrete-time differentiated signal $\dot{x}_1(t)$ which is
additionally low-pass filtered. The low-pass filter (LPF) is
designed as a second-order Butterworth filter with the cutoff
frequency at 200 Hz. The used $x_1(t)$ data are taken from the
closed-loop control experiment (cf. section \ref{sec:3:sub:3}
below) of a 0.5 Hz sinusoidal motion profile, including the
initial transient oscillations of the relative velocity.

\subsection{System identification and PD control tuning}
\label{sec:3:sub:3}

The basic linear model \eqref{eq:2:3:1} of the motion system is
identified from the experimentally collected frequency response
(FR) data $x_1(j\omega) / u(j\omega)$ of the drive. To this end, a
closed-loop identification was performed to keep the drive
position $x_1(t)$ away from the mechanical limits of the operation
range, while allowing for a periodic excitation and motion which
are both required for FR measurement. In the applied control
signal \eqref{eq:3:1:1}, the closed-loop control (here for the
identification purposes only) resulted in
$$
u(t) = k_{id} (r_0 - x(t)) + a \sin \omega_i t.
$$
The proportional feedback gain $k_{id}$ was experimentally tuned
in order to provide the relative zero position at $r_0=6$ mm,
which is approximately the half of the operation range.
\begin{figure}[!h]
\centering
\includegraphics[width=0.95\columnwidth]{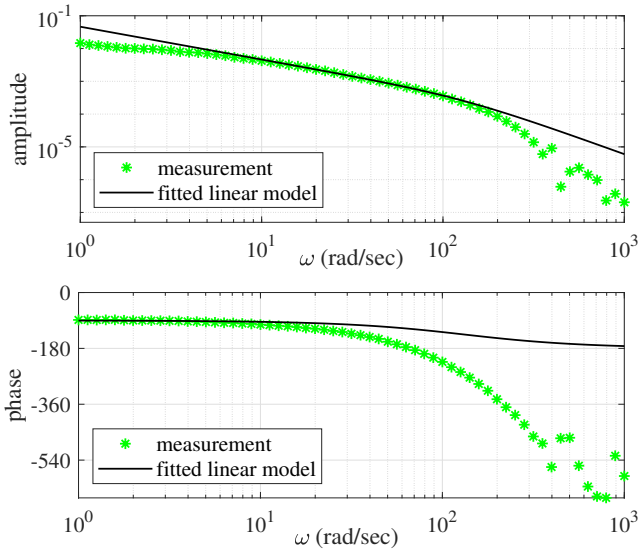}
\caption{Experimentally measured FR versus identified model
\eqref{eq:2:3:1}.} \label{fig:3:3:1}
\end{figure}
The constant amplitude $a$ has also been tuned experimentally, so
as to allow for a sufficient periodic motion with the set of
angular frequencies $\omega_i \in [1, \ldots, 1000]$ rad/sec. The
FR data were collected from the measured steady-sate oscillations
at $\omega_i$, equidistantly distributed on the logarithmic scale,
cf. Figure \ref{fig:3:3:1}. The FR data were used for the
least-squares best-fit of the model \eqref{eq:2:3:1}, while those
low-frequency points were taken out form the FR data set where the
amplitude response violates the $-20$ dB/dec decrease. Recall that
the latter is strictly required for the free integrator of the
system plant, cf. Figure \ref{fig:3:3:1}. The FR identified model
parameters are $K = 0.0463$ and $\tau = 0.0076$. Comparing the
measured and identified frequency characteristics in Figure
\ref{fig:3:3:1} one can recognize that both factors, the gain and
dominant time constant, are sufficiently mapped and, therefore,
identifiable from the collected data set. At the same time, it
becomes apparent that inclusion of an additional (electrical) time
constant of the voice-coil motor, which is known from the
manufacturer's data sheet, would only marginally improve the
amplitude and phase agreement of the model with the measured FR.
From the measured phase response one can recognize much more a
rapid (close to exponential) increase of the phase lag, that can
be attributed to an overall additional time delay $\theta$, i.e.
with the corresponding $\exp(-\theta s)$ transfer characteristics.
This can arise from all sensing, actuating, and power amplifying
elements in the loop. Obviously, this remarkable reduction in the
phase capacity will restrict the overall control gain $\gamma$
and, therefore, the achievable bandwidth.

The reference PD feedback controller, cf. section
\ref{sec:2:sub:4}, is designed based on the identified linear
model \eqref{eq:2:3:1} and the above analysis of the measured FR
of the motion system. Since the differential control part lifts up
the phase characteristics of the open-loop, it is sufficient to
determine the control gain $\gamma$ with respect to the resulted
crossover frequency and the associated phase margin of the
measured FR. The determined $\gamma = 1000$ leads to the crossover
frequency $\omega_c = 46.3$ rad/sec, for which the phase margin
$\pi + \angle \mathrm{FR} (\omega_c) + \angle \mathrm{PD}
(\omega_c)$ of the measured FR characteristics, and further shaped
by the PD controller, results in approximately $\approx 50$ deg,
cf. Figure \ref{fig:3:3:1}. This robust phase margin appears
reasonable for the feedback control design, here for taking into
account the system uncertainties, differentiation of the system
output $x_1(t)$, and non-modeled (to say hidden) residual dynamics
in the control loop.

\subsection{Comparison of PD and OND controllers}
\label{sec:3:sub:4}

Both feedback controllers, the introduced OND \eqref{eq:2:3:5} and
reference PD \eqref{eq:2:4:1}, are experimentally evaluated with
one the same output feedback gain $\gamma = 1000$. Both are
applied (alternately) to the input signal \eqref{eq:3:1:1}. Both
are also sharing the same second-order SMD, designed as in section
\ref{sec:3:sub:2} for use of $y_1(t)$ instead of $x_2(t)$, which
is not available.

\begin{figure}[!h]
\centering
\includegraphics[width=0.9\columnwidth]{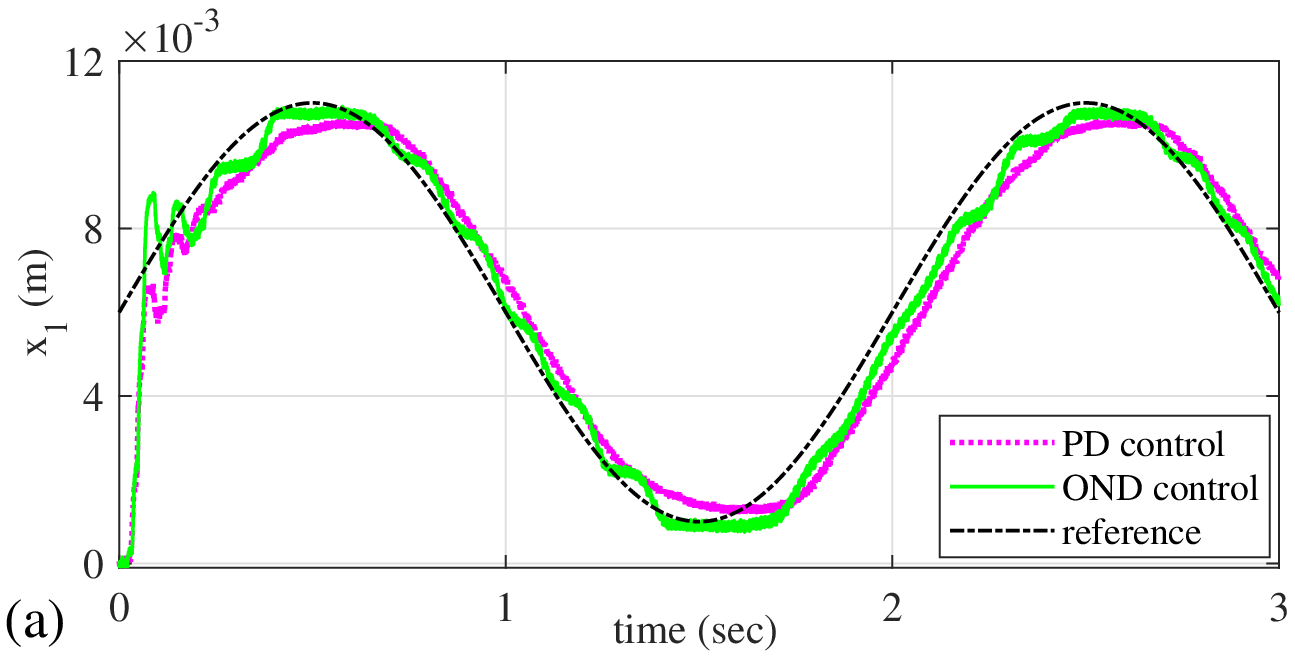}
\includegraphics[width=0.9\columnwidth]{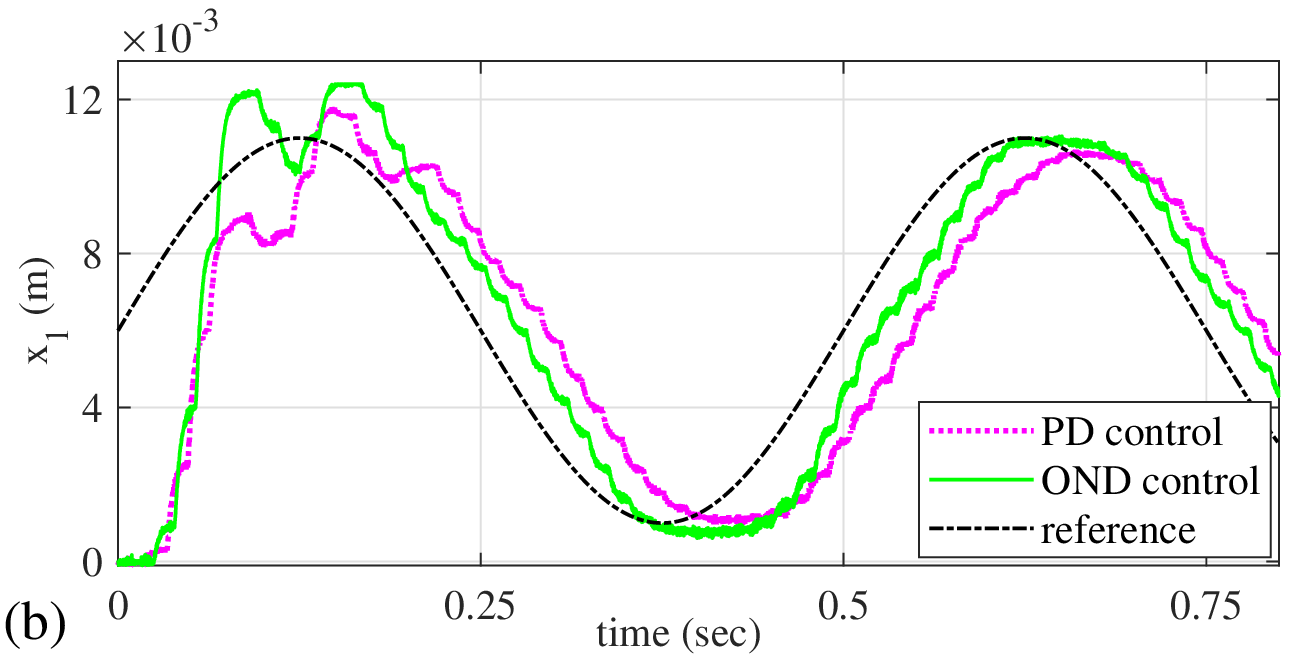}
\caption{Measured position response of OND and PD controls to the
sinusoidal reference trajectory of 0.5Hz in (a) and 2Hz in (b).}
\label{fig:3:4:1}
\end{figure}

First, two sinusoidal reference trajectories are evaluated, one
with 0.5Hz and another one with 2Hz frequency. Both are shown in
Figure \ref{fig:3:4:1} (a) and (b), correspondingly. It is visible
that both controllers have a similar transient response, while the
OND control discloses a lower phase lag and reaches better the
peaks of a periodic trajectory. At the same time, the OND control
tends to a higher initial overshoot, cf. also with state
trajectories in Figure \ref{fig:2:1:1}. This can be explained by a
decreasing damping ratio once $|e_1|$ is growing.

\begin{figure}[!h]
\centering
\includegraphics[width=0.9\columnwidth]{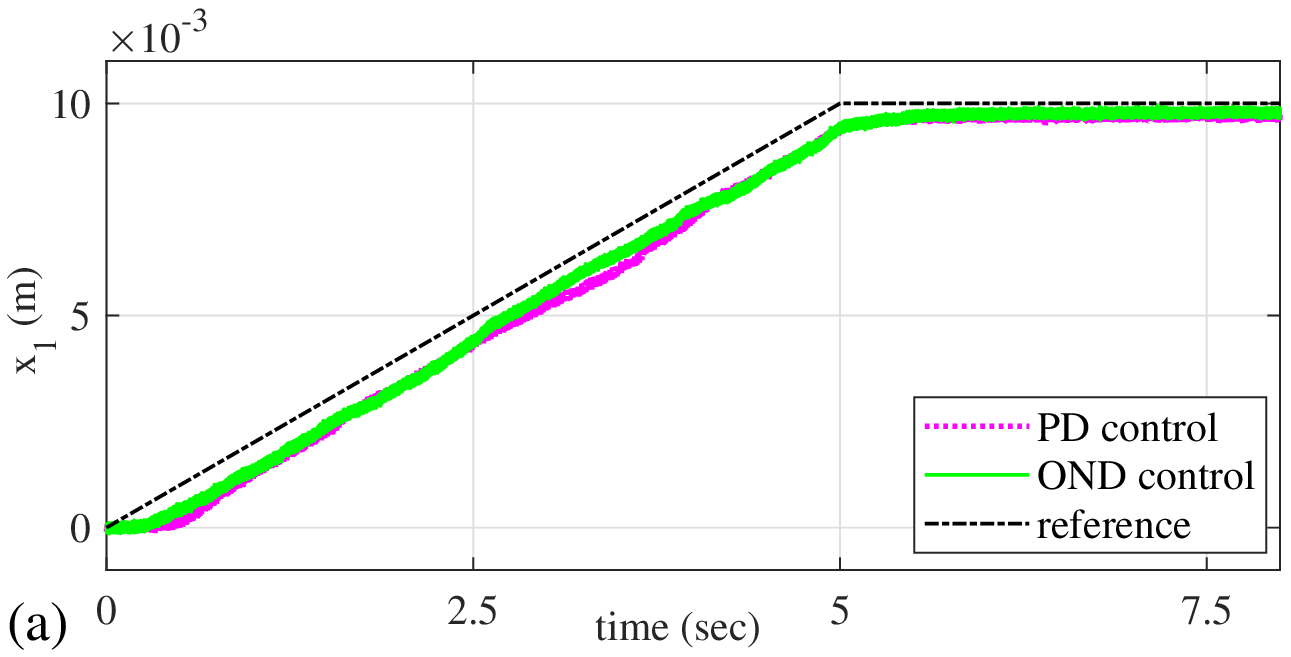}
\includegraphics[width=0.9\columnwidth]{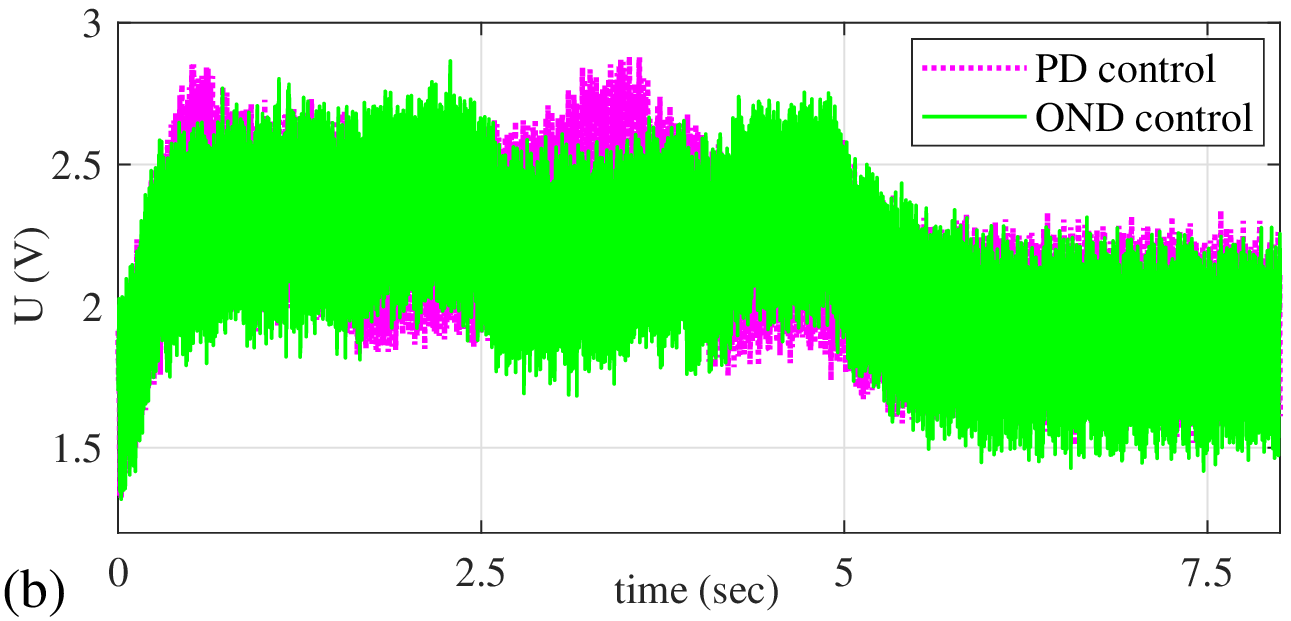}
\includegraphics[width=0.9\columnwidth]{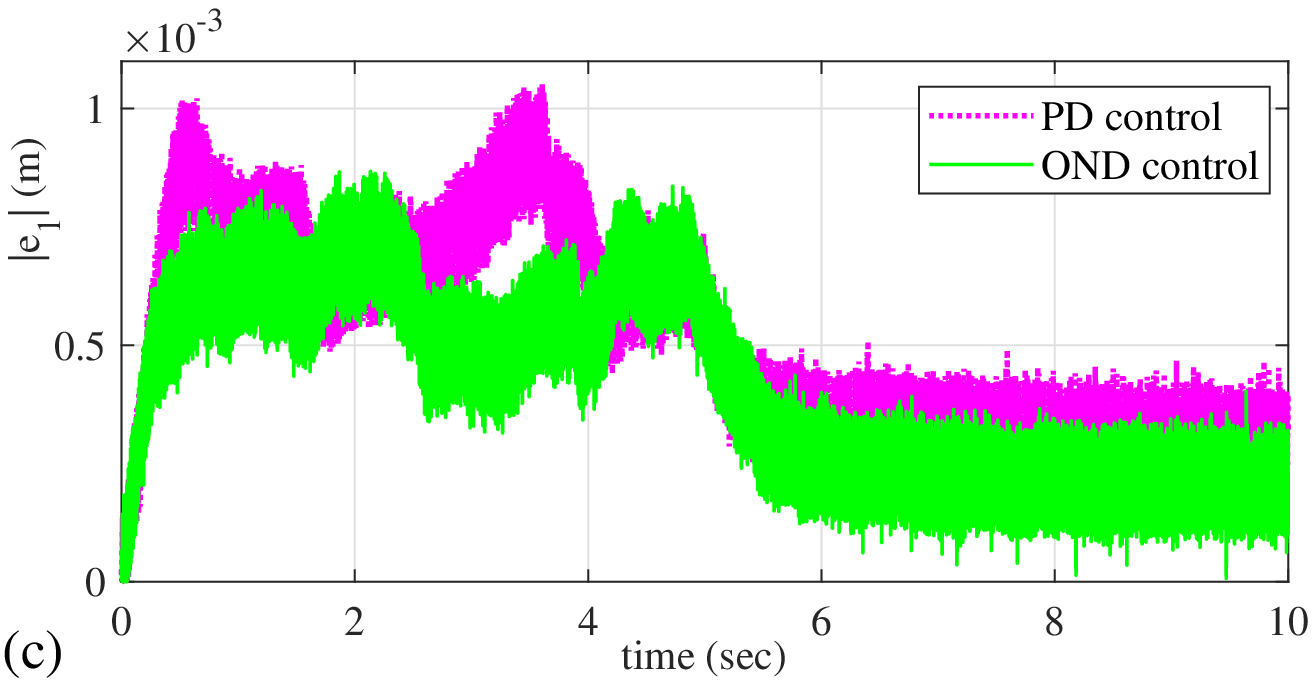}
\caption{Measured position response of OND and PD controls to the
slope reference (a), control value (b), absolute control error
(c).} \label{fig:3:4:2}
\end{figure}

As next, a relatively flat linear slope trajectory is evaluated as
shown in Figure \ref{fig:3:4:2}. This assumed the controlled
motion with a slow constant velocity of $\dot{r} = 0.002$ m/sec.
One can recognize that the OND control is outperforming the PD one
in both following features. (i) It is tracking more uniformly the
reference trajectory, i.e. keeping $e_1(t) \approx \mathrm{const}$
during the steady-state motion. (ii) It is converging closer to
the final reference set value, i.e. having lower $e_1(t)$ at $t
\rightarrow \infty$ in presence of the unavoidable perturbations
$\xi$. Note that the latter can be attributed to e.g. magnetic
force ripples of the void-coil motor and Coulomb friction, see
e.g. \cite{ruderman2015} for details, in the drive. The
corresponding control values (see Figure \ref{fig:3:4:2} (b))
discloses that the OND control is comparable with PD in energy
consumption, and has even slightly lower peaking during the
transient phase and lower average level at steady-state. Recall
that the static bias and jitter are included for both controls,
cf. \eqref{eq:3:1:1}. The control error performance of both
controllers is further visible in more detail in Figure
\ref{fig:3:4:2} (c). Here the same measured data from the slope
reference experiment is used, but with a longer steady-state phase
for better highlighting the $e_1(t)$ performance.

\begin{figure}[!h]
\centering
\includegraphics[width=0.9\columnwidth]{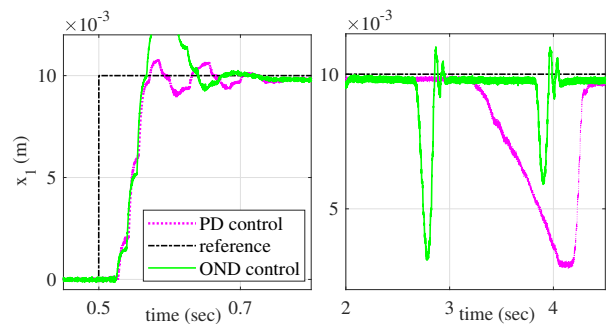}
\caption{Measured position response of OND and PD controls to the
step reference and manually inserted disturbance.}
\label{fig:3:4:3}
\end{figure}

Finally, the ability of disturbance rejection is evaluated for
both controllers. For this purpose, the step reference was used
and, afterwards, the manually inserted disturbance was applied
during the steady-state. The disturbance was induced by pressing
down the moving body of the drive, cf. Figure \ref{fig:3:1:1},
thus inducing an external counteracting force which brings the
motion control away from the constat set reference position $r(t)
= 0.01$ m. The applied disturbance was also released manually,
thus leading to the not repeatable and not equivalent motion
profiles, shown in Figure \ref{fig:3:4:3}. One can recognize that
the OND control behaves more stiff during the step-wise
excitations. In the left zoom-in plot, the step response comes
even beyond the sensor saturation at $x_1=0.012$ m that is,
however, recovered after the transient overshot of the OND
control. In the right zoom-in plot, one can recognize that the OND
control is recovering as fast as the PD control, and comes to the
same residual error level after short transients.

\section{CONCLUSIONS}
\label{sec:4}

In this paper, we have revised the optimal nonlinear damping (OND)
control \cite{ruderman2021a,ruderman2021b} from a more practical
motion control viewpoint. We have extended the admissible plant
dynamics to rather realistic motion systems with damping and gain
factors and bounded matched perturbations. In the theoretical
part, we showed also how an optimal reference PD feedback
controller is parameterized by the same feedback gain as the OND
control, making both controllers well comparable in a fair
benchmark setting. The practical part of the paper introduced and
identified the experimental drive, which is based on the
voice-coil motor and has a relatively high level of the
measurement and process noise. The experimental control evaluation
demonstrated that the proposed OND control is outperforming the PD
reference control in both, the transient response and residual
steady-state error. The OND control proved also to be robust
against the essential external disturbances applied to the
experimental drive. A significantly faster convergence of the OND
control towards zero equilibrium, cf. Figure \ref{fig:2:1:5},
highlights its advantages in case of a more accurate output
position measurement, with the corresponding higher requirements
posed on the motion control system. The following possible issues,
however, have to be taken into account when applying the OND
control. (i) the regularization term $\mu > 0$ cannot be arbitrary
decreased, especially in presence of the measurement noise. For
$|e_1| \rightarrow 0$ and non-zero velocities, the control
dynamics can lead to some residual chattering, owing to the very
small $\mu$ values and alternating sign of $e_2$, cf. with eq.
\eqref{eq:2:1:5}. (ii) the transient peaking of the OND control
can be higher than of the PD control (cf. section
\ref{sec:3:sub:4}), which can provide additional challenges for
certain type of the applications. Despite the above mentioned
shortcomings, it is believed that the OND control represents an
interesting alternative to the conventional PD type controllers,
also with a potential for further developments and extensions,
like for example by an integral control action.

\bibliographystyle{elsarticle-harv}
\bibliography{references}

\end{document}